\documentclass[%
 apl
jmp,%
 amsmath,amssymb,
reprint,%
]{revtex4-1}

\usepackage{natbib}
\usepackage{mathtools}
\usepackage{graphicx}
\usepackage{epstopdf}
\usepackage{enumitem}

\usepackage{dcolumn}
\usepackage{bm}

\begin{document}


\title[
]{Switching probability of all-perpendicular spin valve nanopillars}

\author{M. Tzoufras}
 \email{michail.tzoufras@spintransfer.com}
 \affiliation{Spin Transfer Technologies, Inc., Fremont, California 94538, USA}

\date{\today}

\begin{abstract} 
In all-perpendicular spin valve nanopillars the probability density of the free-layer magnetization is independent of the azimuthal angle and its evolution equation simplifies considerably compared to the general, nonaxisymmetric geometry. Expansion of the time-dependent probability density to Legendre polynomials enables analytical integration of the evolution equation and yields a compact expression for the practically relevant switching probability. This approach is valid when the free layer behaves as a single-domain magnetic particle and it can be readily applied to fitting experimental data.
\end{abstract}

\pacs{?,?,?}

\maketitle

With the recent advent of all-perpendicular Spin Transfer Torque  Magnetoresistive Random-Access Memory (STT-MRAM) \cite{Mangin:2006fk,Ikeda:2010kq}, a significant industrial and academic effort has materialized to better understand  the physics of these devices and to improve their performance. STT-MRAM applications demand high thermal stability, switching speed, and energy efficiency. Assessing STT-MRAM experiments requires knowledge of the theoretically expected performance and there are a number of interrelated models that can be used for this purpose. The simplest and most common approach is to assume that the free layer is a single-domain magnetic particle and simulate the dynamics of the free-layer magnetization with the Landau-Lifshitz-Gilbert (LLG) equation \cite{LandauLifshitz}, augmented by a term for the spin-polarized current \cite{Slonczewski:1996lq,Berger:1996rc}. In this formulation, stochastic fields are employed to introduce finite temperature into the LLG equation  \cite{Brown:1963fk,1367-2630-17-10-103014}. The resulting stochastic LLG equation must be simulated repeatedly for different realizations of the thermal fields to yield good statistics.

Alternatively, the advection-diffusion (Fokker-Planck) equation \cite{Brown:1963fk,1367-2630-17-10-103014} for the probability density of the free-layer magnetization can be discretized and modelled using a standard numerical scheme. Only one copy of such a simulation is needed because the formalism does not involve any stochastic fields; it encompasses finite-temperature effects as a diffusion term in the Fokker-Planck equation. When high fidelity statistics are required, this approach can be faster than the stochastic LLG. Nonetheless, having to represent a distribution on a grid and then update it at each time-step is also computationally intensive. 

Both simulation methods have been applied to study the physics of STT-MRAMs \cite{Li:2004nx,Apalkov:2005qf,Liu2014233,Yxie,Yxie2} but they are too cumbersome  for fitting experimental data, especially when the simulation time is long. To circumvent this limitation, one usually identifies two regimes of distinct physical behaviors \cite{Brown:1963fk,doi:10.1063/1.3532960,LLS-model}: (a) a thermal regime, where the spin current is low and its effect can be approximated by modifying the energy barrier between the two stable states, and (b) a dynamic regime, where the current is high, the spin transfer torque dominates the thermal fields, and the switching process is nearly deterministic. In this latter case the finite-temperature effects are only considered for initializing the system before the current-pulse arrives and are subsequently ignored during the dynamic switching process \cite{Liu2014233}. These approximations can be effective in predicting outlier events, principally in the limits of very low and very high current, but they do not cover the entire range of interesting parameters.

An analytical solution of  the full thermal problem for the general geometry was derived in Refs. \cite{Kalmykov:2013qq,doi:10.1063/1.4754272} by employing the spherical harmonic expansion. Thus, the derivatives in angle were reduced to algebraic expressions and yielded a hierarchy of differential-recurrence relations. These were then recast in matrix form and finally the solution was expressed in terms of Laplace transforms.   

For all-perpendicular spin valves specifically, where the azimuthal angle drops out and the magnetization distribution $\rho$ becomes a function of the polar angle $\theta$ and the time $\tau$ only \cite{Brown:1963fk},  the spherical harmonic expansion reduces to a Legendre polynomial expansion. The Fokker-Planck equation may then be integrated for any current-pulse amplitude and duration, and the solution for  $\rho(\theta,\tau)$ can be expressed in straightforward matrix notation. Obtaining the switching probability from  $\rho(\theta,\tau)$ is a matter of evaluating the dot product between two vectors. Below, I start from the Fokker-Planck equation for the free-layer magnetization probability density $\rho(\theta,\tau)$ of an all-perpendicular spin valve as expressed in Ref. \cite{Butler-FP}:

\begin{equation}\label{FokkerPlanck}
\frac{\partial\rho}{\partial\tau}=-\frac{1}{\sin\theta}\frac{\partial}{\partial\theta}\left[\sin^{2}\theta\left(i-h-\cos\theta\right)\rho-\frac{\sin\theta}{2\Delta}\frac{\partial\rho}{\partial\theta}\right]
\end{equation}
The four normalized quantities in Eq. (\ref{FokkerPlanck}) are: the energy barrier $\Delta=\frac{\mu_{0}H_{k}M_{s}V}{2k_{B}T}$, the external field $h$, the spin current $i$, and the time $\tau$, where  $h,i$, and $\tau$ may be denormalized following Ref. \cite{Butler-FP}:
\begin{align*}
h=\frac{H_{\mathrm{ext}}
}{H_{k}};& \quad H_{k}=\frac{2}{\mu_{0}M_{s}}\left(K_{U}^{B}+\frac{K_{U}^{S}}{t}\right)-N_{zz}M_{s}\\
i=\frac{I}{I_{c0}};&\quad I_{c0}=\frac{2\alpha e}{\eta\hbar}\mu_{0}H_{k}M_{s}V\\
\tau=\frac{t}{\tau_{D}};& \quad  \tau_{D}=\frac{1+\alpha^{2}}{\alpha\gamma\mu_{0}H_{k}}
\end{align*}
$K_{U}^{B}$ is the bulk magnetocrystalline anisotropy,  $K_{U}^{S}/t$ is the surface anisotropy with $t$ the film thickness, and $N_{zz}$ the demagnetizing factor. The parameters $\alpha$ and $\eta$ correspond to Gilbert damping and spin polarization respectively.  

The expansion of the distribution $\rho(\theta,\tau)$ to
Legendre polynomials may be written as:
\begin{equation}\label{LegendreExpansion}
\rho(\theta,\tau)=\sum_{n=0}^{\infty}r_{n}\left(\tau\right)P_{n}\left(\cos\theta\right)
\end{equation}
Where the Legendre polynomials $P_n(x)$ are solutions to Legendre's differential equation $\frac{d}{dx}\left[\left(1-x^{2}\right)\frac{d}{dx}P_{n}\right]+n\left(n+1\right)P_{n}=0
$ and  obey the recurrence relations:
\begin{align}\label{rec1}
\left(n+1\right)P_{n+1}= & \left(2n+1\right)xP_{n}-nP_{n-1}\\
\label{rec2}\frac{x^{2}-1}{n}\frac{d}{dx}P_{n}= & xP_{n}-P_{n-1}
\end{align}

The zeroth order coefficient, $r_0$, is proportional to the number density of the system as can be seen by integrating $\rho$ on the spherical surface: $\int_0^{2\pi}\int_0^{\pi}\rho(\theta,\tau)\sin\theta d\theta d\phi=4\pi r_0$. Similarly, the first order term corresponds to the expected magnetization $\langle m_z(\tau)\rangle = \int_0^{2\pi}\int_0^{\pi}\rho(\theta,\tau)\cos\theta\sin\theta d\theta d\phi = (4\pi/3) r_1(\tau)$. 

 Substitution of the expansion (\ref{LegendreExpansion}) into the Fokker-Planck equation (\ref{FokkerPlanck}) and application of the recurrence relations (\ref{rec1})-(\ref{rec2})  yields:
\begin{equation}\label{FokkerPlanckSH}
\frac{\partial\rho}{\partial\tau}=\sum_{n=0}^{\infty}\sum_{k=-2}^{2}r_{n}a_{n+k,n}P_{n+k}
\end{equation}
with the coefficients $ a_{\imath,\jmath}$   defined as:
\begin{align}
a_{n-2,n}= &-\frac{n\left(n-1\right)\left(n-2\right)}{\left(2n+1\right)\left(2n-1\right)}\label{Penta1}
 \\
a_{n-1,n}= &\left(i-h\right)\frac{n\left(n-1\right)}{2n+1}\label{Penta2}
\\
a_{n,n}= &-n\left(n+1\right)\left[\frac{1}{2\Delta}-\frac{1}{\left(2n+3\right)\left(2n-1\right)}\right]\label{Penta3}
\\
a_{n+1,n}= & -\left(i-h\right)\frac{\left(n+1\right)\left(n+2\right)}{2n+1}\label{Penta4}
\\
a_{n+2,n}= &\frac{\left(n+1\right)\left(n+2\right)\left(n+3\right)}{\left(2n+1\right)\left(2n+3\right)}\label{Penta5}
\end{align}
Rearranging Eq. (\ref{FokkerPlanckSH}) and using the orthogonality of the Legendre polynomials enables integration of the Fokker-Planck equation:
\begin{equation}\label{Matexponential}
\frac{\partial\mathbf{r}}{\partial\tau}=\bm{\mathit{A}}\mathbf{r}\Rightarrow\mathbf{r}(\tau)=e^{\bm{\mathit{A}}\tau}\mathbf{r}(0)
\end{equation}
where $\mathbf{r}=(r_n)$ is the vector of the expansion coefficients and  $\bm{\mathit{A}}=\left(a_{\imath,\jmath}\right)$ is the pentadiagonal matrix with elements in Eqs. (\ref{Penta1})-(\ref{Penta5}). All elements in the first row of $\bm{\mathit{A}}$ are identically equal to $0$, i.e. $a_{0,n}\equiv 0,\forall n$. This ensures that---since there are no sources or sinks of spin valves in Eq. (\ref{FokkerPlanck})---the number density $4\pi r_0$ is constant.

 In deriving Eq. (\ref{Matexponential}), the spin current $i$ and the external field $h$ were assumed to be independent of time. In general, when $\frac{\partial}{\partial\tau}  (i-h)\neq 0$, the matrix $\bm{\mathit{A}}$ is time-dependent with $\bm{\mathit{A}}(\tau_1)\bm{\mathit{A}}(\tau_2) \neq \bm{\mathit{A}}(\tau_2)\bm{\mathit{A}}(\tau_1)$. The solution for $\mathbf{r}(\tau)$ should then be expressed using the Magnus expansion  \cite{CPA:CPA3160070404}  instead of the matrix exponential from Eq. (\ref{Matexponential}). Below, only the case of constant driver, $\frac{\partial}{\partial\tau}  (i-h)= 0$, is discussed.

With the probability density in terms of the expansion coefficients $\mathbf{r}(\tau)$ from Eq. (\ref{Matexponential}), the probability that the magnetization is in the upper hemisphere,  $\mathcal{P}(m_z>0)$, can be calculated as:
\begin{equation}
\mathcal{P}(m_z>0)=
 2\pi \,\mathbf{r}^\intercal\mathbf{s} \label{notSwitched}
\end{equation}
where $\mathbf{r}^\intercal\mathbf{s} $ is the dot product between the vectors $\mathbf{r}$ and $\mathbf{s} = \left(s_n\right)$ with elements given by (see Ref. \cite{Byerly}): 
\begin{equation}
s_n = \intop_{0}^{1}P_{n}(x)\,\mathrm{d}x=\begin{cases}
1 & n=0\\
\frac{(-1)^{\frac{n-1}{2}} n!!}{n\left(n+1\right)\left(n-1\right)!!} & n\textrm{ odd}\\
0 &\mkern-20mu 
n\textrm{ even, }n\neq0
\end{cases}
\end{equation}

\begin{figure}[thbp]
\includegraphics[width=9.1cm]{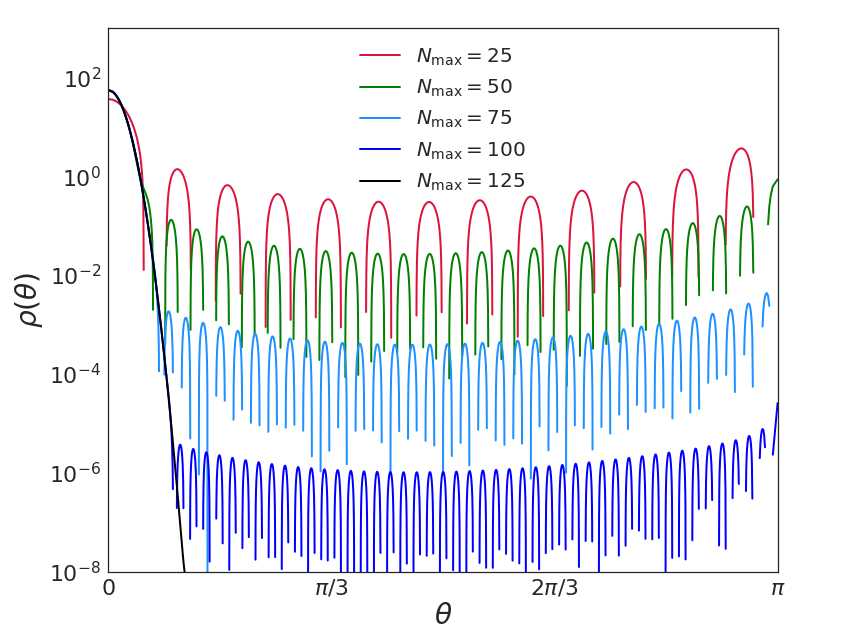}
\caption{\label{fig:Nmax} 
Analytical solutions of Eq. (\ref{Matexponential}) for a system with  $\Delta = 60$, $i-h=-2$, $\tau=10$. An initial $r_{n}(0)=0, \forall n >0$, was chosen, with normalization $r_0=(4\pi)^{-1}$. The different lines correspond to different cutoff, $N_{\max}$, in the Legendre polynomial expansion. The figure shows that for these parameters the probability density can be calculated to $10$ orders of magnitude accuracy with $N_{\max}\sim125$.} 
\end{figure}

Typically, one is interested in the probability that  a device with a given energy barrier, $\Delta$, at some initial $\mathbf{r}(0)$, switches orientation for certain external field and current-pulse amplitude and duration. Using Eqs. (\ref{Penta1})-(\ref{notSwitched}) the required switching probability can be calculated by performing the following four operations: 
\begin{enumerate}[label=(\alph*)]
\item construction of the matrix $\bm{\mathit{A}}$,
\item matrix exponentiation: $e^{\bm{\mathit{A}}\tau}$, 
\item matrix-vector multiplication:  $\mathbf{r}(\tau)=e^{\bm{\mathit{A}}\tau}
\mathbf{r}(0)$,  
\item dot product: $ \mathcal{P}(m_z>0)=2\pi\, \mathbf{r}^{\intercal}\mathbf{s}$.
\end{enumerate}
Of the four operations listed above, only matrix exponentiation can be of non-trivial computational cost, and this only happens when the number of terms $N_{\max}$ that are retained in the Legendre polynomial expansion is large.

However, for problems of practical interest $N_{\max}\sim100$ usually suffices and the resulting $100\times100$ matrix $\bm{\mathit{A}}$ can be immediately exponentiated by standard linear algebra software packages. The reason behind the rapid convergence of the expansion rests with  Eqs. (\ref{Penta1})-(\ref{Penta5}): while the off-diagonal terms scale as $O(n)$ the diagonal ones scale as $O(n^2)$, therefore $\frac{\partial}{\partial\tau} r_{n\gg1}\simeq -\frac{n\left(n+1\right)}{2\Delta}r_{n\gg1}\Rightarrow r_{n\gg1} \sim e^{-\frac{n\left(n+1\right)}{2\Delta}\tau}\simeq 0$. In other words, since $\Delta$ is finite for $T>0\mathrm{K}$, highly peaked components of the distribution tend to diffuse away. In Fig. \ref{fig:Nmax}, I present a number of distributions generated by applying Eq. (\ref{Matexponential}), where I vary $N_{\max}$.  This figure confirms that a relatively modest increase in $N_{\max}$ leads to a significant improvement in accuracy for the probability density $\rho(\theta,\tau)$.

In the limit  $n\ll N_{\max}$, the second term on the right hand side of Eq. (\ref{Penta3}), which derives from the anisotropy inherent in the magnetic material,  becomes comparable with the first and counteracts diffusion, that is, it prevents $r_{n>0}$ from vanishing. To see this, one can consider a system without a net external driver  ($i-h=0$) after a very long time ($\tau\rightarrow\infty$). In this scenario, it is known that the magnetization orientation is not completely randomized, but it lies along the anisotropy axis, and therefore  $r_{n>0}\left(\tau\rightarrow\infty\right) \neq 0$. Still, for sufficiently large values of $n$ the effect of magnetic anisotropy on $r_n$ is overwhelmed by thermal diffusion and $r_n$ diminishes as discussed above. The critical value $n=n_c$ for which the diffusion term overtakes the magnetic anisotropy and $a_{n,n}$ becomes negative is: $n_c=\sqrt{\Delta/2+1}-1/2$. Hence, higher $\Delta$ is associated with higher $n_c$, and increasing the thermal stability produces magnetization distributions with sharper peaks on the anisotropy axis that require more terms in the Legendre polynomial expansion.
   
The off-diagonal terms $a_{n\pm1,n}$ are proportional to $i-h$. When these terms vanish, both directions on the anisotropy axis are equivalent; without a spin current or an external field there is nothing to break the symmetry of the bistable system. If one starts from a  symmetric system without a net external driver and lets the system relax, the final distribution will also be symmetric.

\begin{figure}[thbp]
\includegraphics[width=9.1cm]{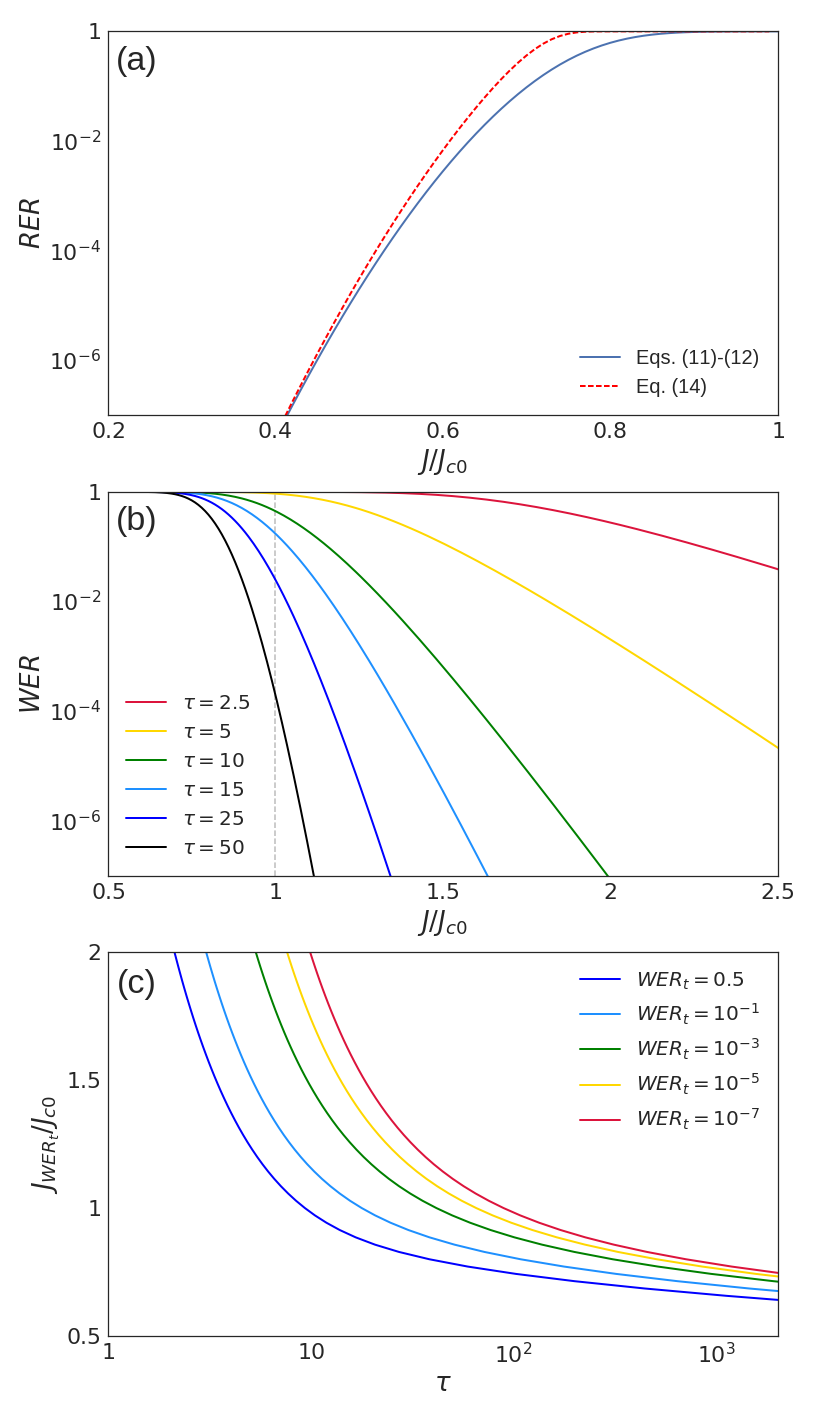}
\caption{\label{fig:Fig2abc} 
Examples using Eqs. (\ref{Matexponential})-(\ref{notSwitched}). For all of the cases shown in this figure: $\Delta=60$, $h=0$, $\tau_D=2\mathrm{ns}$, and $N_{\max}=125$.
(a) The Read Error Rate [$R\!E\!R \equiv \mathcal{P}(m_z<0) $] for a $100\mathrm{ns}$ pulse as calculated from Eqs. (\ref{Matexponential})-(\ref{notSwitched}) (solid blue line) and from an approximation in Ref. \cite{Liu2014233}  (Eq. (\ref{dexp}), dashed red line). (b) The Write Error Rate [$W\!E\!R\equiv \mathcal{P}(m_z>0) $] as a function of the normalized current $J/J_{c0}$ for pulse-lengths from $5\mathrm{ns}$ to $100\mathrm{ns}$. (c) The spin current  $J_{W\!E\!R_t}/J_{c0}$ required to achieve a specific error rate target, $W\!E\!R_t$, as a function of the pulse-length $\tau$.} 
\end{figure}

In practice, one usually attempts repeated switches of a device, or an ensemble of devices, for a range of pulse amplitudes and durations and counts the number of failed switches for each set of parameters. This yields the error rate as a function of amplitude or pulse-length. The  same approach can be taken using  Eqs. (\ref{Matexponential})-(\ref{notSwitched}): Starting from the trivial case $r_{n>0} = 0$, one generates a distribution as in Fig. \ref{fig:Nmax} and then lets it relax by multiplying it with a suitable matrix exponential following Eq. (\ref{Matexponential}). This relaxed distribution, $\mathbf{r}(0)$, becomes the initial condition from which new distributions are generated by applying Eq. (\ref{Matexponential}). The error rate is calculated by substituting each new distribution, $\mathbf{r}(\tau)$, into Eq. (\ref{notSwitched}).

For example, the ``read disturbance"  method for evaluating thermal stability via the parameter $\Delta$ involves: attempting to switch a device with low current amplitude $i$, generating the error rate as a function of $i$, and fitting the measured data with an approximate expression such as:
\begin{equation}\label{dexp}
\mathcal{P}_{\vert i\vert\ll1}(m_z>0)=e^{-f_0\tau\exp\left[-\Delta (1-i)^{\beta^\prime}\right]} 
\end{equation}
In this expression $f_0\simeq1GHz$ is the attempt frequency and the exponent $\beta^\prime$ was calculated analytically in Ref. \cite{Liu2014233} for all-perpendicular geometry: $\beta^\prime = 2$. It is assumed that $h=0$. The inner exponential derives from the N\'{e}el-Arrhenius model \cite{Neel} and the outer from the approximation that the switching time is much longer than $1/f_0$, which also implies $\vert i-h\vert\ll 1$. Unfortunately, it can be  difficult to determine the experimental $\mathcal{P}_{\vert i-h\vert\ll1}(m_z>0)$ because very few switches happen under conditions of weak disturbance and a massive number of measurements needs to be taken to ensure adequate statistics. It is more convenient to obtain reliable data in the intermediate range, where $ \vert i-h\vert\sim1$, which is  beyond the range of validity of Eq. (\ref{dexp}). In Fig. \ref{fig:Fig2abc}(a), the Read Error Rate [$R\!E\!R \equiv \mathcal{P}(m_z<0) $] is shown for pulse-length $100\mathrm{ns}$ from two separate calculations: the solid blue line derives from Eqs.  (\ref{Matexponential})-(\ref{notSwitched}) and the dashed red line from Eq. (\ref{dexp}). Good agreement between the two approaches is found in the limit $\vert i-h\vert\ll 1$. However, Eq. (\ref{dexp}) tends to overestimate the $R\!E\!R $ when $\vert i-h\vert\sim1$.

In Fig. \ref{fig:Fig2abc}(b) the Write Error Rate [$W\!E\!R\equiv \mathcal{P}(m_z>0) $] is shown as a function of the normalized spin current $i=J/J_{c0}$ for various pulse-lengths. The parameter $\tau_D=2\mathrm{ns}$ is chosen to facilitate comparison with Fig. 13 from Ref. \cite{0022-3727-46-7-074001}, which was generated using Eq. (3.15) from Ref. \cite{Butler-FP}. Although the expressions in Refs. \cite{0022-3727-46-7-074001,Butler-FP} recover the correct asymptotic behavior of the $W\!E\!R(J/J_{c0})$ in the limit $J\gg J_{c0}$, they fail in the region $J\lesssim J_{c0}$, especially for relatively long pulses, $\tau>10$. The lines shown in Fig. \ref{fig:Fig2abc}(b) were calculated by directly applying Eqs. (\ref{Matexponential})-(\ref{notSwitched}). 

Another useful plot is that of the switching current $J_{W\!E\!R_t}$ in terms of the pulse-length, where $J_{W\!E\!R_t}$ is defined as the spin current density required to achieve a specific $W\!E\!R$ target, $\mathcal{P}(m_z > 0)=W\!E\!R_t$. In Fig. \ref{fig:Fig2abc}(c), $J_{W\!E\!R_t}$ is shown as a function of the pulse-length for $\!W\!E\!R_t=0.5$ as well as for smaller  $W\!E\!R_t$ values. These lines were calculated by numerically solving the transcendental equation $2\pi \left[e^{\bm{\mathit{A}}\tau}
\mathbf{r}(0)\right]^\intercal\mathbf{s}=W\!E\!R_t$.

In summary, from the Fokker-Planck equation for the probability density of the free-layer magnetization of an all-perpendicular spin valve nanopillar, $\rho(\theta,\tau)$,  a compact analytical solution, Eq. (\ref{Matexponential}), was derived by employing the expansion of $\rho(\theta,\tau)$ to Legendre polynomials. From this expression the expected switching probability can be straightforwardly calculated using Eq. (\ref{notSwitched}) under the assumption that the free layer acts as a single-domain magnetic particle.

\bibliographystyle{unsrt}

\end{document}